**Multi Linear Regression applied to Communications systems Analysis.**

Written by:
MSC. Federico Rodas Bajaña
MSC. Luis Hernan Montoya Lara
MSC. Manolo Paredes
Ph.D. Elena Gimenez de Ory
Ph.D. Luis Manuel Diaz Angulo

**Abstract**

*This paper develops a propagation model of electromagnetic signals emitted at frequencies of 20 and 40 MHz for the Ecuadorian jungle. It is expected that the results obtained at the end of this research will be applied to produce a complete coverage map for wireless communications technologies, which will optimize the radio spectrum in operations carried out by the Armed Forces in the Ecuadorian border jungle. The final expression found is an adjustment function that relates the Receiving Power ($P_{RX}$) to factors that determine the geometry of the Fresnell Zone (Connectivity). The resulting model of the research improves the discrepancy between the simulated power ($P_{RX}$) in commercial software and a sample of measured wireless transmissions in situ. The analysis was based on the results and methodology presented by Longley-Rice. It was determined the non-normality of the discrepancy between the losses ($L_{LR}$) calculated by Longley-Rice Model (LMR) and the data obtained in the field, It was added correction coefficients on the expression of LMR. Subsequently, the mathematical expression was linearized to implement multivariate linear adjustment techniques. Alternative formulations to the Linear Regression model were sought and their goodness of fit was compared; all these techniques are introduced theoretically. To conclude, an analysis of the error of the found model is made. Mathematical modeling software such as MATLAB and SPSS were used for the formulation and numerical analysis, whose algorithms are also introduced. Finally, we propose future lines of research that allow a global understanding of the behavior of telecommunications technologies under hostile environments.*

1. **Introduction**

Basic mobile phone systems, wireless data networks, digital television or radio links communication of data, voice and video to large regions. Currently, maintaining efficient control systems in certain regions is a problem both for civil society and for the Army to carry out permanent military operations in these sectors due to conflicts related to drug control patrols and guerrillas especially. In this context, it is necessary to determine a specific propagation model adjusted to this type of characteristics, therefore, the eastern region has been selected as a reference scenario to test the Longley-Rice propagation model, since it allows to establish a specific analysis for rural sectors and in a scenario extremely complex, where no previous studies have been developed for such purposes. The main objective of this work is to solve and design a regression model that allows to predict with an error less than 5% the Power ($P_{RX}$) under the above conditions including the variability of all the factors described in this paragraph. The study sample obtained for the formulation of the function adjustment of propagation obtained in Cevallos et al. (2009), is a data set of five variables: reception powers ($P_{RX}$) in dB, distance between transmitter and receiver (d) in meters, transmission power ($P_{TX}$) in dB, height above sea level (h) in meters and frequencies (f) in MHz, for wireless transmissions in the 20- 40 MHz. The sample size is 256 x 5, i.e. 256 vector data formed by the variables described were taken. The methodology used for the experimental and sampling process was exposed by two works carried out to obtain the same statistical treatment using the model LMR, and commercial software, but on Mediterranean environments and in data transmissions at frequencies used in WLAN technologies, higher than 2 GHz ( Kasampalis, S., Lazaridis, P., Zaharis, Z., & Cosmas, J., 2013) and (Kasampalis, S., Lazaridis, P., Zaharis, Z., & Cosmas, J., 2014). After the completion of this study, it is recommended to perform a complete analysis of the frequency response of the vegetation of the Ecuadorian rainforest, and the generalization of the final result for the entire spectrum of frequencies used in mobile communications technologies.

In Section 3, the theoretical formulation of the Longley-Rice Model (LMR), widely used for wireless network simulation and popularized due to its implementation in commercial software such as Radio Mobile and Sirenet, which are the most widely used thanks to its ease of use and its graphical interface, example in Figure 1. On the other hand, given the conditions of great weather variability, commercial software for link survey do not predict with a high degree of certainty the power $P_{RX}$ in jungle areas. Thus, the Ecuadorian Army proposes the development of software that has the same features but that introduces corrections to reduce the error evidenced in the calculations. In Section 4, the mathematical techniques needed to develop a linear fit model as well as simple and multi-variant will be summarized. These tools include Multiple Linear Regression, and Least Squares Adjustment, as well as computational tools for the numerical treatment of the problem in question. In Section 4, the general structure of the propagation model described in Section 3 is used and adapted to a functional that can be worked using the techniques set out. Section 6, and 8 particular emphasis is placed on the process of filtering the data and refining the sample. It also describes the calculation problems that appeared during the process. The data collection process is described in detail in Section 6, which also describes the technical characteristics of the equipment used for data collection, and the geographical and meteorological conditions of the measurement site. Mathematical and statistical analysis is discussed in depth in Section 7. Here we compare the results obtained after applying different techniques of multiple linear adjustment and choose the model with the greatest goodness of fit. In addition, validation tests of the final results are included.

## 2 Methodology

Given that the study variable is the Receiving Power in dB ($P_{RX}$). The Longest link for VHF signal transmissions, is less than 20 Km., with frequencies included in the 20 to 40 MHz. band, and low, medium and high Transmission Powers ($P_{TX}$) corresponding to 1W, 5W and 20W respectively, and that after data collection was obtained as a result a sample of size 1000 for each of the above factors. It is important to note that data must be gathered during both dry and wet weather. In addition, we want to measure the oscillation caused by environmental changes, however this work is reserved for future studies. Data also has to be filtered using thermal noise levels given the working frequency of the transmission. The dependence of these two variables is well described theoretically under the name White Noise (Nyquist H., 1928), and whose basic principle determines that for the bands to which we are going to refer, the power values for this noise would be between [-100, -144] dB, however, in the study of the sample size only were less than -120 dB, so it was necessary to apply the same filtering criterion to delete data between [-100, -120] dB. After filtering and with the purged sample, the actual data will be compared with the predicted by the Longley Rice Model (LMR). If normality exists in the resulting residue vector, it would indicate that the error is random and that this randomness corresponds to a characteristic of the sample and not to external factors such as weather conditions and signal dissipation due to the vegetation of the area. If the waste vector does not have normal function, then the error could be found an adjustment function that allows to discriminate the influence of each of the factors mentioned.

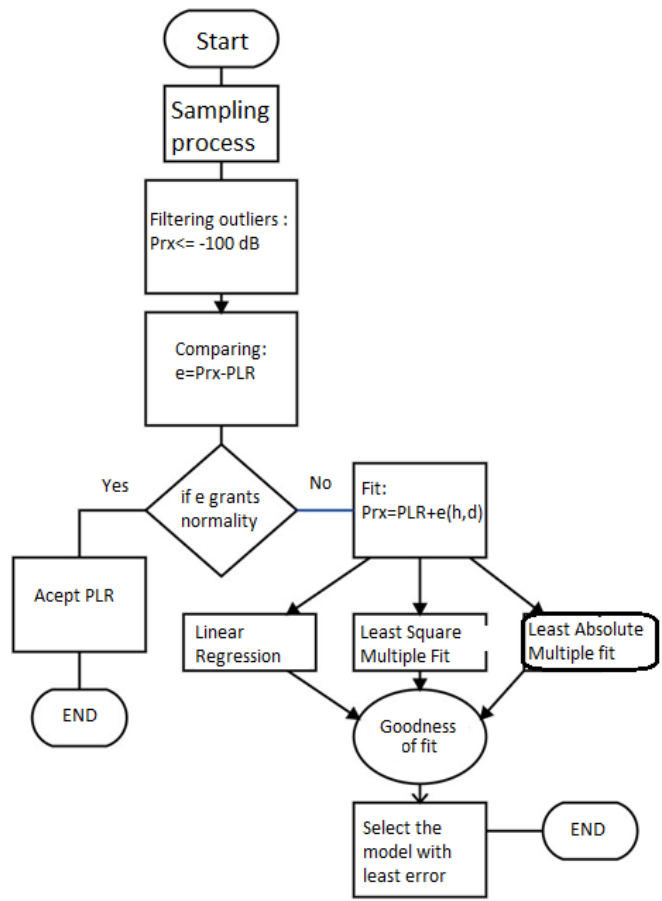

Figure 1: Analysis Process

Given the nature of our study, we propose the use of adjustment processes for the determination of such a non-deterministic model. This whole process will be discussed in quite detail in the development of the work and can be summarized in Figure 1. For the analysis of the results, mathematical and statistical analysis packages such as Matlab and SPSS will be used, with which regression and Least Squares Regression (LSR) analysis is proposed. It is important that after regression analysis, the behavior of the random error between the LMR and the experimental data collected in the field is observed. It is proposed that this study will be complemented in the future with Montecarlo techniques, which facilitate the simulation of the experimental process through the use of computers; in this way the process would be automated to reach the sweep of all the electromagnetic spectrum used for wireless technologies.

## 3. Propagation models

Propagation models are responsible for predicting the attenuation of the power of a signal, the analysis of which contemplates a signal received at a certain distance, as well as changes in the intensity of the signal received, based on measurements in a specific geographical area for a given system or technology. The propagation model to be used for the present project is the model LMR, therefore, it will be deepened in its formulation and theorization. ITU-R defines the propagation of free space as the propagation of an electromagnetic wave in a homogeneous and isotropic ideal dielectric medium that can be considered infinite in all directions (International Telecommunication Union, 2007). The free space model indicates the signal power assuming that the radio frequency channel is free of any obstacle.

The Equation 1 is well known as "The Friis Transmission Equation" (Friis H.,1946) indicates the ratio for the received and transmitted power between two separate antennas at a distance d:

$$P_r = (P_t G_t G_r \lambda^2)/((4\pi)^2 d^2) \qquad (1)$$

### 3.1 Okomura Hata model vs Longley Rice

The Longley-Rice model was proposed for frequencies between 20 MHz and 20 GHz. This model uses statistical parameters which depends on the variables of each environment. The variation of the signal depends on atmospheric changes, topography and free space. These variations are described with the help of statistical estimates .

On the other hand, The Longley-Rice model also called ITM (Irregular Terrain Model), Irregular Terrain Model was created in 1968, used for the study of TV signal propagations and models distant and near obstacles" (Isaiah, 2015). The Longley-Rice model can determine the losses in transmission links over long distances by the obstacle diffraction phenomenon using Fresnel-Kirchoff's "Knife Edge" model.

The Longley-Rice propagation model estimates the average power transmitted based on the information on the geometry of the terrain and the refractive characteristics of the troposphere" (Salazar, 2005). The attenuation value is always a continuous function of the distance traveled. This model reaches a distance of 1 to 2000 km and is applied to several communication systems, one of them is point-to-point communication. You can determine the trajectory specific parameters in the frequency range of HF to EHF with ranges from 20 MHz to 20 GHz. Used to make area predictions, predicts the average attenuation of a radio signal based on distance and signal variability (Rice, 2000). The Longley Rice method establishes a loss of propagation by means of the geometry of the terrain in the receiver. Mainly uses the 2-ray terrestrial reflection model.

**Limits**

The operating range of this system is as follows:
- 20 MHz ≤ frequency ≤ 20 GHz.
- 1 Km ≤ distance ≤ 2000 Km.
- 0,5 m ≤ Antenna height Rx ≤ 300.0 m.

Environmental parameters are as follows: Polarization: The Longley-Rice model specifies that the polarization of the antennas either horizontal or vertical should be the same at both points. Refractivity: Refractivity determines the amount of "bending" of radio waves when an wave collides to a boundary between two media and the waves do not penetrate the second medium. The propagation model determines the amount of curvature that will have the radio waves , by default the value of 1,333 is used for normal atmospheric conditions.

Terrain irregularity Δh models uses linear interpolation for a range of interest and then determines the interval Δh that characterizes the size of the terrain. Table 1 presents a range of values and qualitative descriptions of the terrain.

| Type deTerrain | $\Delta h$ |
|---|---|
| Water or flat Surface | 0-10 |
| Flat surface | 10-20 |
| Lightly irregular surface | 40-60 |
| Hills | 80-150 |
| Mountains | 200-500 |

Table 1: Types of terrain, (Rice, 2000)

**Refractive index Ns**: The value of surface refractivity lays between 250 to 400 units N (depends on the curvature of the earth). The surface refractivity value is 301 units N, which is used for the Longley-Rice propagation model. Furthermore, surface refractivity is given by the Equation 2 (W. L. Patterson, 1990):

$$N_S = 179.3 \ln((1 - 1/k)/0.046665) \qquad (2)$$

K is the effective value of curvature as 1,333 for the Earth.

**Attenuation:** The distance and elevation related to the horizon allow to calculate the loss per transmission relative to free space .The model divides the total transmission loss into "basic transmission loss in free space" and the reference attenuation relative to free space. For short ranges the propagation model is not particularly sensitive to changes in the value of Ns refractivity, in which they have a definite effect on transmission losses.

**Permittivity (ε):** Also called dielectric constant, it describes the electric field (E) that affects and is affected by the medium. Table 2 shows the values that correspond to the typical values of the permittivity.

**Conductivity**: Conductivity is a physical property that allows measuring the energy conduction capacity of materials, measured in Siemens per meter with typical values indicated in Table 3. Conductivity and permittivity establish the electrical behavior of materials that determine the induced current density and the load density, provided that an E electric field is applied.

| N | Weather | Ns |
|---|---|---|
| 1 | Equatorial | 360 |
| 2 | Continental Subtropical | 320 |
| 3 | Marine Subtropical | 370 |
| 4 | Desert | 280 |
| 5 | Template | 301 |
| 6 | Marine template, upon earth | 320 |
| 7 | Marine template, upon wáter | 350 |

Table 2: Types of weather, (Rice,2000)

| Type | Permittivity $[C/m^2]$ | Conductivity $[S/m]$ |
|---|---|---|
| Medium Earth | 15 | 0.005 |
| Poor Earth | 4 | 0.001 |
| Rich Earth | 25 | 0.02 |
| Fresh water | 81 | 0.010 |
| Sea water | 81 | 5 |

Table 3: Permittivity y conductivity, (Rice,2000)

**Variability**: The variability model presents through a mathematical demonstration a random variable, which establishes environmental parameters, and chooses an event by means of a probability measurement based on time and location. The variability parameters used in the Longley Rice propagation model correspond to time, location and situation. "Time variability is found in changes due to atmospheric phenomena." (Kingdom and Albert, 2017). The variability by location intervenes in the terrain profiles and finally by situation is considered data outside the stipulated, the effects of which are not of interest for this analysis. Each variability is expressed in percentages of 0.1% to 99.9%.

### 3.2 Calculation of attenuation

The ITU (International Telecommunication Union) is the specialized agency of the United Nations responsible for regulating telecommunications at the international level; it has validated and disseminated a number of international technical standards developed by the Radio Industry (ITU-R), which are useful to understand the behavior of the propagation of radio signals and other related phenomena. The established ITU method for forecasting coverage is a method that predicts the applicable E value for broadcasting, mobile links and fixed communications. The following are the recommendations issued by ITU-R that were used in the development of this project: "BO.794 (03/92) Techniques to minimize the effect of the influence of rain on the connection link relating to the general characteristics of the satellite broadcasting service systems. Recommends that, when introducing digital sound broadcasting services with terrestrial transmitters for reception in vehicles, portable and fixed on kilometer, heptametric and decametric wave bands, the digital system used offers high quality monophonic or stereo sound, a compromise solution between amplitude coverage and quality of service for a given emission power". ITU-R Recommendation P.526-11 states: "The methods for calculating field intensities in diffraction propagation pathways, which may correspond to the surface of a spherical Earth or to uneven terrain with different types of obstacles, consider that taking into account the loss of edge diffraction Geometric parameters are used grouped in a dimensionless value designated as "Cascaded knife edge method" which provides the Eq. 3.

$$v = h\sqrt{(2/\lambda) \cdot (1/d1 + 1/d2)} \qquad (3)$$

- *h* is the height of the top of the obstacle on the line connecting the two ends of the path.
- *d1* and *d2* are distances from the two ends of the path to the top of the obstacle.

Recommendation ITU-R P.833-2 states that "Vegetation attenuation should be assessed at frequencies between 30 MHz and 60 GHz. The model allows to determine the excess loss experienced by the signal when passing through vegetation. In practice, once the signal crosses the vegetation receives contributions due to the propagation of both vegetation and diffraction that occurs around it. The maximum attenuation value, limited by the dispersion of the surface wave, depends on the type and density of the vegetation, as well as the radiation diagram of the terminal antenna within the vegetation and the vertical distance between the antenna and the highest point of the vegetation."

### 3.3 Okomura-Hata model

Modeling losses in free space involves transforming Power applying logarithms on both side of the definition of "Gaining Power". These definitions are depicted in the Eq. 4.

$$P_{RX} = A \cdot P_{TX} \qquad (4)$$

$$\log(P_{RX}) = \log(A) + \log(P_{TX}) \qquad (5)$$

(with L annotated pointing to $\log(A)$)

Where A is the gain and L are the losses in open space, Kasampalis et al (2014)

Okomura Hata defines losses in open space for distances less than 20 Km. as the Equation 6 shows:

$$\begin{aligned}L_H[dB] = {} & 69{,}55 + 26{,}16\log_{10}(f_{MHz}) - 13{,}82\log_{10}(h_1) \ldots \\ & - 1{,}1\log_{10}(f) + \ldots h_2 + (1{,}56\log_{10}(f) - 0{,}8) \ldots \\ & + (44{,}9 - 6{,}55\log_{10}(d_{km}))\end{aligned} \qquad (6)$$

Where $h_1$ represents the height of the base station, and $h_2$ is the height of mobile station. The Eq. 6 provides a functional to be fit with the model for electromagnetic waves propagation in open spaces. This functional will be studied further in this work.

4. Numerical and statistical analysis

**4.1 Multivariate Linear regression**
A linear regression model should provide the Eq. 7. Besides , to avoid the influence of the sign of Residuals during the process, it is necessary to square power the residuals vector as it is shown in the Eq. 9 as follows:

$$y_i = \beta_0 + \sum_j^n \beta_j x_{ij} + \varphi_i \qquad (7)$$

$$\begin{bmatrix} y_1 \\ y_2 \\ \vdots \\ y_n \end{bmatrix} = \begin{bmatrix} 1 & x_{11} & \ldots & x_{k1} \\ & & & \\ & & & \\ & & & \end{bmatrix} \begin{bmatrix} \beta_0 \\ \beta_1 \\ \vdots \\ \beta_k \end{bmatrix} \qquad (8)$$

$$\sum_i^n \varphi_i^2 = \sum_i^n \left( y_i - \beta_0 - \sum_j^k \beta_j x_{ij} \right)^2 \qquad (9)$$

**Linear Least Square Method (LES).** Given the residuals, it is necessary to minimize their influence over the final result. Then it is necessary to get a model granting that the residual vector tend to zero which is easy to be accomplished by getting the partial derivative of the residual with respect to the coefficients. The process is well described by the Eq.10, Eq.11, Eq. 12, (M. Bremer, 2012), as follows.

$$\min \left( \sum_i^n \left( y_i - \beta_0 - \sum_j^k \beta_j x_{ij} \right)^2 \right) \quad (10)$$

$$\frac{\partial \varphi_i(a,b)}{\partial \beta_0} = -2 \left\{ \sum_{i=1}^n \left( y_i - \beta_0 - \sum_{j=1}^k \beta_j x_{ij} \right)^2 \right\} = 0 \quad (11)$$

$$\frac{\partial \varphi_i(a,b)}{\partial \beta_0} = -2 \left\{ \sum_{i=1}^n \left( y_i - \beta_0 - \sum_{j=1}^k \beta_j x_{ij} \right)^2 \right\} x_i = 0 \quad (12)$$

Let´s transform the system in to a matricial operation:

$$Y = \begin{bmatrix} y_1 \\ y_2 \\ \vdots \\ y_n \end{bmatrix}, \quad \beta = \begin{bmatrix} \beta_1 \\ \beta_2 \\ \vdots \\ \beta_n \end{bmatrix}, \quad \varphi = \begin{bmatrix} \varphi_1 \\ \varphi_2 \\ \vdots \\ \varphi_n \end{bmatrix} \quad (13)$$

$$X = \begin{bmatrix} 1 & x_{11} & x_{12} & \cdots & x_{1k} \\ 1 & x_{21} & x_{22} & \cdots & x_{2k} \\ \vdots & \vdots & \vdots & & \vdots \\ 1 & x_{n1} & x_{n2} & \cdots & x_{nk} \end{bmatrix} \quad (14)$$

$$then, Y = X\beta$$

Now, multiplying both sides for the transpose X', we get

$$X'Y = X'X\beta$$

given a matrix H defined as:

$$H = (X'X)^{-1}X'$$

Then, the matrix of the vector of coefficients $\beta$ is determined as:

$$\beta = H.Y \quad (15)$$

Therefore, this axiomatic algorithm leaves the final expression for the fitting process concluded.

### 4.2 Plane of linear regression

In practice, point clouds do not consist of a small value of them, rather most adjustment involves many values. It is necessary to find the line that represents a better fit for all these points and this is the basis of the least squares method. Using this method, you can calculate the slope value and the ordered value at the origin of the best estimate and to obtain them the following consideration must be taken into account: contributions due to the propagation of both vegetation and diffraction that occurs around it.

$$L(f, d, h) = A \log f + B \log Di + C.h + D \qquad (16)$$

Goodness of fit
- Standard error of the estimate:

$$S_e = \sqrt{\frac{\sum(y - \hat{y})^2}{n - k - 1}} \qquad (17)$$

- The coefficient of multiple determination:

$$R^2 = 1 - \frac{\sum(y - \hat{y})^2}{\sum(y - \bar{y})^2} \qquad (18)$$

### 4.3 Checkout process

To analyze data using linear regression, part of the process involves verifying that the data wanted to analyze might actually be analyzed using linear regression. The process is suited, if data checks six assumptions that are necessary for obtaining a validated linear regression.

**Data continuity:** Your two variables must be measured at the continuous level (i.e., they are interval or relationship variables). Examples of continuous variables include review time (measured in hours), intelligence (measured using the IQ score), test performance (measured from 0 to 100), weight (measured in kg), and so on. It is important to clarify that this process can be carried out to non-continuous variables, using other procedures, but that will not be dealt with in this paper.

**Multiplicity:** There must be a linear relationship between the dependent variable and the independent ones. It can be said that there is linearity if the partial correlation coefficients $r_{ij}$ that measures the relationship between the dependent variable with each of the independent variables are high. The correlation matrix R is defined as:

$$R = \{r_{ij}\} \qquad (19)$$

$$r_{ij} = S_{xy}/(S_x S_y) \qquad (20)$$

Where S is the standard deviation of the variable x or y, if the case is y depending on the subscript, and $S_{xy}$ is the covariance between x and y.

Another way to demonstrate multicolliniality is by using the Variance Independence Factor (IVF), which must be less than the tolerance (Tol) which is defined as:

$$Tol = 1 - R_i^2 \qquad (21)$$

,such that i represents the number of independent variable

**Outliers**: An outlier is an observed data point that has a dependent variable value that is very different from the value predicted by the regression equation. As such, an outlier will be a point in a scatter diagram that is (vertically) far away from the regression plane indicating that it has a large residue. The problem with outliers is that they can have a negative effect on regression analysis (for example, reducing the adjustment of the regression equation) that is used to predict the value of the dependent variable based on the independent variable. This will change the result produced by SPSS and reduce the predictive accuracy of your results.

**Principle of Parsimony**: This means that the model should be explained with as few variables as possible. For this purpose, the Durbin-Watson test is used which verifies independence between the variables which is a simple test to run with SPSS, the null hypothesis (Ho) suggests that the residues between both variables are random therefore independent, to accept the null hypothesis the Durbin-Watson statistic must take values between 1.5 and 2. 5.

**Homoscedasticity:** Your data should show homoscedasticity, which is where variations along the best-fit line remain similar as you move along the line. To check homoscedasticity, it is enough to observe that the dispersion around the regression line should not vary much between each independent variable, but more technically you can use the Levene Test which has as H or the equality of variance for each independent variable. You can safely ignore this assumption if you have (approximately) the same sample size for each group. However, if the group sizes are (markedly) different, then you need to make sure that your data meets the homogeneity of the variations using the Levene test, however the results of the Levene test are identical to the Analysis of Variance (ANOVA), so the process and mathematical model will be explained later, such as general rule, we conclude that variances are not equal if the significance (sig) is less than 0.05.

**Residue normality:** Finally, you should verify that the residues (errors) of the regression line are distributed normally. Two common methods to verify this assumption include using a histogram (with a normal curve superimposed) or a normal P-P graph, or using the Kolmogorov-Izmir Test (TKE), however, the use of (TKE) is preferable for analytical rigor.

### 4.5 Fitting using quasi linear Least Absolute Residuals (LAR)

Errors must follow a normal distribution, but still there are extreme values called outliers that we already introduced in the previous topic. Since the disadvantage of least-squares adjustment is its sensitivity to outliers. Outliers have a great influence on adjustment because squaring debris increases the effects of these extreme data points. To minimize the influence of outliers, you can adjust your data using robust least squares regression. The toolbox provides these two robust regression methods: Minimum Absolute Waste (LAR): The LAR method finds a curve that minimizes the absolute difference of waste, rather than square differences. Therefore, extreme values have less influence on the fit.

$$\sum_{i=1}^{n} |\varphi_i| = \sum_{i=1}^{n} \left| y - \beta_0 - \sum_{j=1}^{k} \beta_j x_{ij} \right| \qquad (22)$$

### 5. Mathematical software and statistical packages.

Errors must follow a normal distribution, but still there are extreme values called outliers that we already introduced in the previous topic. Since the disadvantage of least-squares adjustment is its sensitivity to outliers. Outliers have a great influence on adjustment because squaring debris increases the effects of these extreme data points. To minimize the influence of outliers, you can adjust your data using robust least squares regression. The toolbox provides these two robust regression methods:

#### 5.1 Matlab

Matlab is a commercial scientific programming language, whose Kernell is written in C for numerical calculation and Java for graphical representation. Its greatest functionality is the verstaility it has for the handling of matrices and vectors that allows to perform computations using command window. For the case of application of Computational Statistics, matlab presents a fairly complete library that covers a wide range of functionalities. For linear adjustment, the polyfit (), and polyval () functions will be used.

An example of simple linear regression can be seen in Figure 2, where its output per screen is also observed using the functions listed above.

```
> > c = [82 98 87 40 116 113 111 83 85 126 106 117];
> > h = [42 46 39 37 65 88 86 56 62 92 54 81];
```

> > p = polyfit (c, h,1) p = 0.6628 -1.9592
> > x = linspace (min (c), max (c));
> > y = polyval (p, x);
> > plot (c, h, "o ")
> > hold on; plot (x, y," r ")
> > grid on

Table 4: Example of linear regression using Matlab

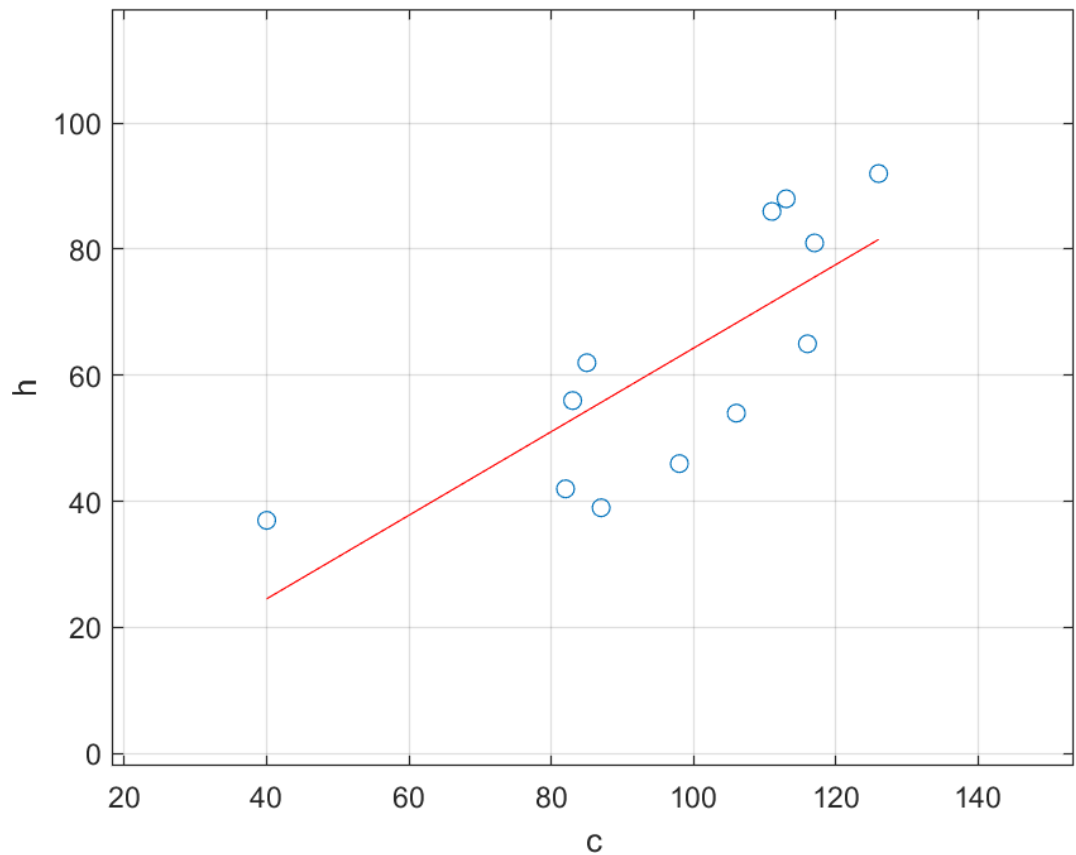

Figure 2: Output of linear regression example using Matlab.

It also has a user environment that allows to perform simple curve adjustment as multiple, as well as other features such as splines adjustment, polynomial adjustment, and interpolation. The application window , as well as an example, is shown in Figure 3.

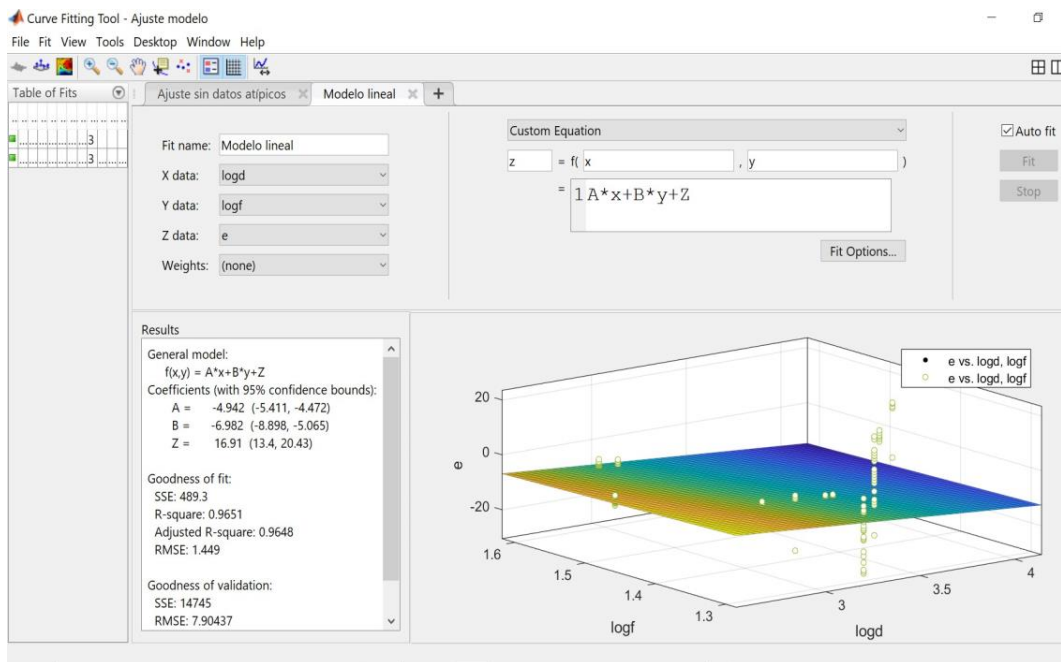

Figure 3: Least Squares Adjustment Example using CFTOOL.

## 5.2 SPSS

SPSS is a platform with a user interface organized in cells similar to Excel but that specializes in the use of advanced statistical tools, which gives it a great robustness for its application in Big Data and Multivariate Statistics. Unlike Matlab, its greatest strength is not the output per console, but rather, the use of a rather intuitive user interaction, without losing the power of statistical calculation, such as it allows to perform multiple non-linear regression, a feature that gives it some advantage over Matlab that does not allow perform any type of nonlinear adjustment with more than two independent variables. The environment is based in Windows OP, as it follows in the next figure.

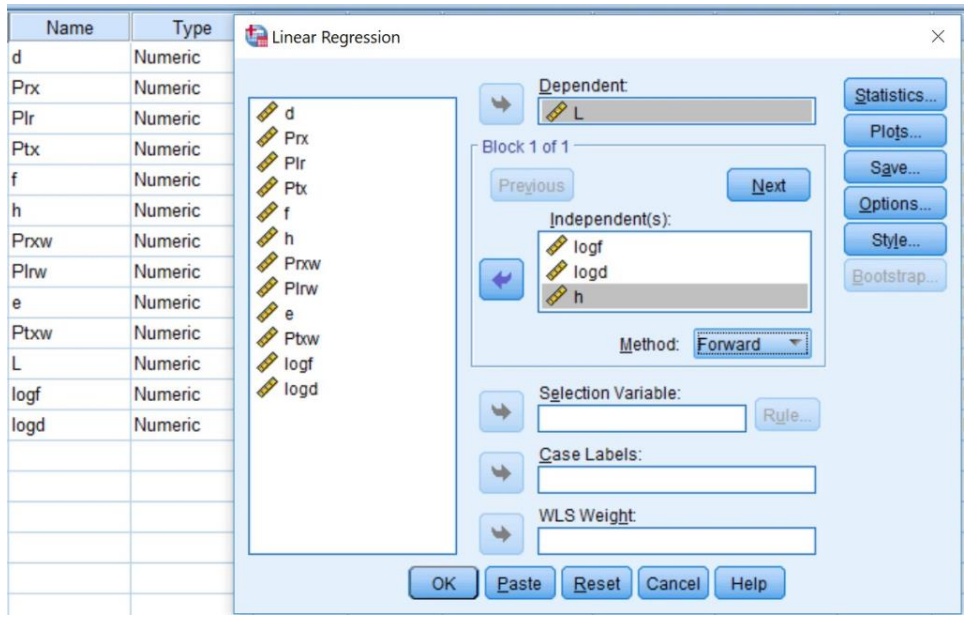

Figure 4: Linear Adjustment Using SPSS

## 6. Sampling and data collection

First of all, the study location of the area should be analyzed to establish a theoretical estimate of the range calculation, frequency band, climatic characteristics of the zone and technical specifications of transmission and reception equipment, within these is antenna gain, receiver sensitivity; error rate and transmission and reception power. The installation of equipment must be carried out and the results obtained must be analyzed and then compared with radio software that allows us to estimate the coverage of each radio link, as well as the possible levels of interference and losses.

The equipment selected to establish the different radio links must meet the requirements set out in Table 4, which presents characteristics of the equipment used as transmitter and receiver. These characteristics are those used by military equipment used by the troops in the intended locations.

For the measurement of reception power, a Spectra Analyzer was used, equipment that allows to perform essential RF measurements. It has a vector network analyzer, spectrum analyzer, and vector voltmeter. This equipment allows to measure and monitor the reception power of the various broadcasting frequencies, has an operating range of 2 MHz to 6 GHz, and an impedance of 50 Ohms.

## 7. Simulation of theoretical data using the Radio Mobile Simulator

It is a radio propagation simulation program developed by Roger Coude in 1998, allows predicting behavior of radio systems (Pellejero, 2009). Uses digital terrain elevation data to generate path profile between an emitter and receiver. Radio Mobile software uses the Longley-Rice propagation model also called ITM (Irregular Terrain

Model), this radio software allows to simulate the coverage area in a communication system (Tomasi, 2003), and downloads elevation maps such as Global 30 Arc-Second Elevation (GTOPO30), Digital Terrain Elevation Data (DTED) and SRTM; models developed by NASA for free use. Applications such as radio system performance prediction, site assessment, GPS interface among others can be highlighted. In addition, it has didactic tools for analysis of coverage of simulated links. The software will allow us to simulate the coverage and give us the theoretical values of our sample as seen in Figure 5.

| Parameters | Measures |
|---|---|
| Frequency band | HF |
| Center frequency | 25 MHz |
| Transmission power | 1,5,20 Watts |
| Modulation | AM |
| Tx antenna height | 5 |
| Rx antenna height | 2.5 |
| Polarization | Vertical |
| Climate | Equatorial |

Table 4: Technical Characteristics to use during the sampling process

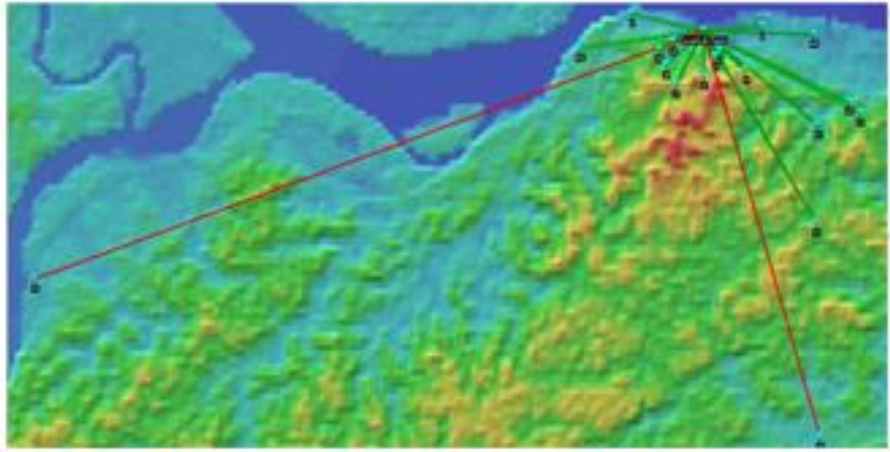

Figure 5: Simulation of theoretical values according to the Longley Rice model, using Radio Mobile software

### 8. Normality of the residuals vector

Given the significance of the constants based on comparing tolerance with the VIF statistic, the use of the variable representing height in the first and second model is discarded, which physically keeps some logic. Let´s define the error (e) of the Longley-Rice model as the difference in value between the value in the receiver power (Prx) and the power forecast delivered by the Longley-Rice model (Plr), such that

$$e_{LR} = \log(P_{RX}) - \log(P_{LR}) \quad (23)$$

To be able to make an adjustment it is important to show that a dependent variable is not random and one way to check this is to see that it does not comply normally, and this is the graph of a CDF for a $N(\nu = 0, \sigma^2)$, as seen in Fig. 6 we see that the adjustment can do using the distribution function (CDF) the same that should fail in the adjustment with close to zero and order 1e15., in this way it would be evidenced the non-normality of a normal CDF is poor, also using the Kolmogorov test in MATLAB with the function kstest (), which receives as parameter the residue vector giving the significance of the statistic said residue vector. Then the null hypothesis is rejected, and in this way we assume that there is no randomness in the data of the error, this would give way for an adjustment function to be performed.

As the model LMR is a linearized by pulling logarithms on both sides of we could say that our adjustment model can be expressed as follows:

$$\log(P_{RX}) = \log(P_{LR}) \pm e_{LR} \quad (24)$$

Hopefully, since the model is a function of the "f", "d", "h", "L.". Then by transmission of the error we expect to have a reception function of type:

$$\log(P_{RX}) = A \log(P_{LR}) + B \log(f) + C \log(d) + Dh + E \qquad (25)$$

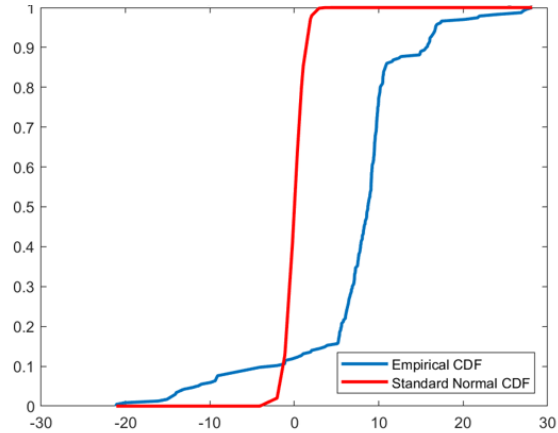

Figure 6: CDF for the Longley Rice Model

**8.1 Filtering Outliers**

The linear least squares model allows minimization of the sum of the differences squares. In the case of a linear model there is no difference between the regression model and the least squares model, so the mathematical management consists in a few words is to guarantee the six assumptions, which are: continuity, linearity, multicollinearity, homoscedasticity, parsimony and normality of the residuals.

First of all, to determine continuity suffice to say that variables take data with decimal places to ensure this point. Second, the linearity is performed analyzing the degree of correlation of the variables, if we observe the correlation matrix in Table 6 high values of the r statistic are observed, both for $\log(P_{RX}) - \log(P_{LR}) - \log(d)$ and $\log(P_{RX}) - \log(d)$, so it is observed that there is a linear relationship between the variables above except the height (h), so the generation of a linear model of these characteristics is recommended but taking into account the index of IVF higher than the tolerance to avoid multicollinearity, the height variable is deviated, resulting in a canonical model.

SPSS presents high functionality to perform an initial analysis, but due to its difficulty in programming scripts, Matlab has a better performance for the development of software that facilitates the development of a script that allows the automation of the process for future studies in other positions of the spectrum. If we recall that the central objective of the research was to obtain an adjustment model with an error of less than 5% of the deviation, it is observed that this target would be more than covered.

In addition to improving the homoscedasticity of the model, it is important to filter the data to avoid anomalous data. An initial analysis can be done by fixing the significance of the F statistic of the ANOVA analysis, which should be very close to zero.

It is concluded that the sample presents homoscedasticity, however, to improve the response of our model, anomalous data are eliminated. The criterion used for the deletion of such data was a technical criterion, since receiving powers below 120 dB are considered noise. The comparison of both processes can be seen in Figures 7 and 8. Then the SSE (Sum of Squeare Error) factor that is the sum of the squared of the waste decreases ostensibly, this is an indication of the improvement of the error of each model.

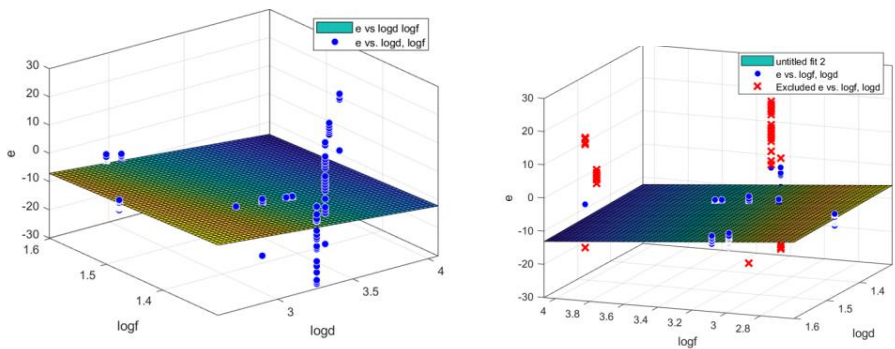

Figure 7-8: Outliers filtering. Multivariate linear fitting model for $P_{RX}[dB] = A \log(d) + B \log(f) + Z$

### 8.2 Normality of the Residuals Vector (l-r)

In order to have a conclusive model, reproducibility of the same is important; this means that the final formulation can reproduce its response with other data that are within the domain of the adjustment function. In order to apply this principle, about 25 data have to be taken which were not taken into account in the calculation of the final refined model. Then, the fitted function must be applied on the test sample using the model LMR and the final model obtained in the previous section. Finally, these divided standard deviation are compared for the mean of each vector, and if the final model is correct, the error obtained by the calculated model remains less than the error obtained by the LMR model.

To complete the evaluation of the model, it is important to determine whether the resulting residue vector meets the normal condition. Then a Kolmogorov test is performed with the residue vector. Comparing the significance of the model (p factor) with that of the Longley Rice Model, the larger to 0.05 is better, in this way it is observed that ostensibly the error generated by the non-deterministic model improves the behavior compared to that of Longley Rice, however it continues to reject the null hypothesis of the Kolmogorov-Smirnoff test, which implies not normality. At the end of the process we will see that the CDF of the residual vector better fits to a normal CDF function than previously observed for the residue vector of the Longley Rice model as it is shown in the Figure 9.

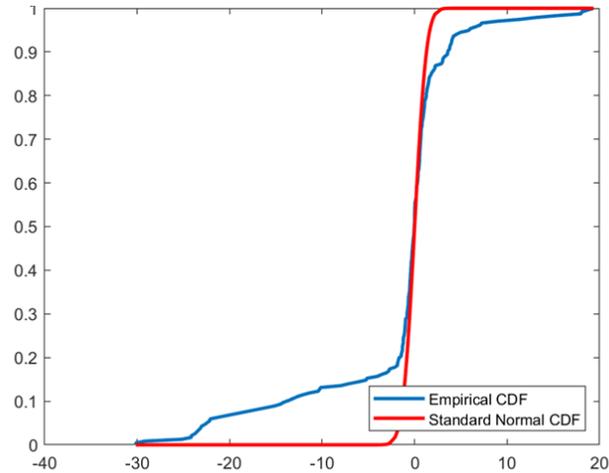

Figure 9: CDF for the obtained model

**Conclusions**

1. The adjustment model for the calculation of power P RX that best behaves on the sample is: log ($P_{RX}$) = A log (d) - B log (f) + log ($P_{LR}$), it seeks representing at least over the 95% of the total variance. It is evidenced that the Quasi-Robust LES techniques improves the goodness of the results compared to the linear regression model, that conclusion is provided by comparing the R of correlation which explains the variance percentage explained by the Fit Function. It is finally expressed as: L = A-B h- C log (d) - D log (f)

2. The general error was calculated through the coefficient of variation, and it must be less than 5%.

3. Since the differences vector of the Longley Rice model ($e_{LR}$) does not meet the normality characteristics as observed in Figure 6, it is clear that after adjustment there is still a dependence of the residue vector resulting from the (LMR) to the variables already taken into account in the process. However, there is a considerable improvement in the assessment of this parameter with respect to (LMR). In addition, the lack of normality of the CFD of the differences vector is seen as a result of a loss of information in the sampling process that does not allow the error to behave clearly as white noise. This lack of randomness may be due data could be biased, since the sample was taken only in the dry season of the Ecuadorian rainforest. It is advisable to plan a new study including a rainy season measurement schedule to eliminate such behavior from CFD.

4. After filtering the sample, the model increases its representativeness as well as decrees the standardized error. This is seen by comparing the Tabs tab: Set-linear-using and tab: Set-linear-con. There is also an improvement in the correlation factor R, which although the change is minimal is representative; a fact that must be taken into account for future studies at different working frequencies.

5.  5. It is observed that there is a variation of the power multiplier coefficient $P_{LR}$ so it could be understood as a correction to the model (LMR). Since this model already uses meteorological factors to describe the variation of the Fresnell Area, it is recommended to conduct a study of the variation of reception power due to the weather of the area, as well as a study of attenuations due to native vegetation.